\documentclass{jetpl}
\usepackage{cite}
\twocolumn
\lat
\title{On differential equation on four-point correlation function in the Conformal Toda Field Theory}

\rtitle{On differential equation on four-point correlation function \dots}

\sodtitle{On differential equation on four-point correlation function in the Conformal Toda Field Theory}

\author{V.\,A.\,Fateev$^{+*}$,
A.\,V.\,Litvinov$^{+}$ \thanks{e-mail: litvinov@itp.ac.ru}}

\rauthor{V.\,A.\,Fateev , A.\,V.\,Litvinov}

\sodauthor{Fateev, Litvinov}

\address{$^+$L.D.Landau Institute for Theoretical Physics RAS, 
142432 Chernogolovka, Russia\\~\\
$^*$Laboratoire de Physique Th\'eorique et Astroparticules, Universit\'e
Montpelier II, Pl.E. Bataillon, 34095 Montpelier, France}

\dates{\today}{*}

\abstract{The properties of completely degenerate fields in the Conformal Toda Field Theory are studied. It is shown that a generic four-point correlation function that contains only one such field does not satisfy ordinary differential equation in contrast to the Liouville Field Theory. Some additional assumptions for other fields are required. Under these assumptions we write such a differential  equation and solve it explicitly. We use the fusion properties of the operator algebra to derive a special set of three-point correlation function. The result agrees with the semiclassical calculations.}
\begin{document}
\maketitle
There are several motivations to study the Conformal Toda Field Theory. This theory has a non-trivial geometric formulation 
\cite{Gervais:1993yh} and hence plays a significant role in 
the quantization of non-critical strings  with extended symmetry \cite{West:1993np}. It also provides an example of a theory with higher spin symmetry and hence has its own interest.
The algebra of generators of this symmetry($W$ algebra) is closely related with rather general class of integrable systems. It can be derived by quantization of the second Hamiltonian structure of the  generalized KdV-type equations associated with algebras Lie 
\cite{Fateev:1991aw,Belavin:1990jf}. This symmetry manifests 
itself in rational Conformal Field Theories (CFT) which describe the critical behavior of many interesting statistical systems like for example $Z_n$ Ising models (parafermionic CFT \cite{Fateev:1985mm}), tricritical Ising and $Z_3$ Potts models, Ashkin-Teller models etc.  Known for many years and being applied to many interesting problems in contemporary mathematics and physics the $W$ symmetry still stay rather mysterious and need further detailed study. The Conformal Toda Field Theory with real coupling constant is  irrational CFT which has a simple Lagrangian formulation and posses this symmetry. In this letter we find some special set  of three-point correlation function of the exponential fields in this theory, but general formula is not known for us at present. 

The action of the Conformal Toda Field Theory has the form
\begin{equation}
\mathcal{A}=\int d^{2}x\left( \frac{1}{8\pi}(\partial\varphi)^{2}+\mu \sum_{k=1}^{r}e^{b(e_k,\varphi)}\right), 
\end{equation}
where $e_k$ are the simple roots of Lie algebra $\mathcal{G}$ and 
$(e_k,\varphi)$ denotes the scalar product of the roots with $r$ component scalar field $\varphi=(\varphi_1,\dots,\varphi_r)$. We consider the case $\mathcal{G}=sl(n)$, $r=n-1.$ The conserved holomorphic $W$ currents $W_{j}$ which form closed $W$ algebra in this case have spins $j=2,...,n$ and can be expressed in terms of $\varphi$ by the relation\footnote{This relation is known in the theory of 
integrable equations as a Miura transformation\cite{Fateev:1991aw}.}
\begin{equation}\label{WCurrents}
\prod\limits_{i=0}^{n-1}(q\partial+(h_{n-i},\partial\varphi)
)=\sum_{k=0}^{n}W_{n-k}(z)(q\partial)^{k}, 
\end{equation}
where $q=b+1/b$  and $h_{k}$ are the weights of the first fundamental representation $\pi _{1}$ of $sl(n)$ with the highest weight $\omega_{1}:h_{1}=\omega_{1},h_{k}=\omega_{1}-e_{1}-\dots-%
e_{k-1}$. In particular, $W_{0}=1,W_{1}=0$ and
\begin{equation*}
W_{2}=T(z)=-\frac{1}{2}(\partial\varphi)^{2}+(Q,\partial^2\varphi) 
\end{equation*}
is the stress-energy tensor. Here $Q=(b+1/b)\rho$ with $\rho$ being the Weyl vector (half of the sum of all positive roots).
Primary fields of the $W$ algebra are exponential fields
\begin{equation*}
V_{\alpha}=e^{(\alpha,\varphi)}.
\end{equation*}
The main term of the operator product expansion (OPE) of these fields with the currents $W_k(z)$ defines the quantum numbers $w^{(k)}(\alpha)$
\begin{equation}
W_k(z)V_{\alpha}(z^{'})=\frac{w^{(k)}(\alpha)V_{\alpha}(z^{'})}{(z-z^{'})^k}+\dots.
\end{equation}
They are known explicitly  and are symmetric under the action of the Weyl group $\mathcal{W}$ of the Lie algebra $sl(n)$ \cite{Fateev:1987zh}
\begin{equation}\label{WeilSymmetry}
w^{(k)}(\alpha)=w^{(k)}_s(\alpha)\equiv w^{(k)}(Q+s(\alpha-Q)),\quad s\in\mathcal{W}.
\end{equation}
In particular,
\begin{equation*}
w^{(2)}(\alpha)=\Delta(\alpha)=\frac{(\alpha,2Q-\alpha)}{2}
\end{equation*}
is the conformal dimension of the field $V_{\alpha}$. The equation \eqref{WeilSymmetry} means that the fields  connected via the  action of the Weyl group $\mathcal{W}$ should coincide up to a multiplicative constant
\begin{equation}
V_{Q+s(\alpha-Q)}=R_s(\alpha)V_{\alpha}.
\end{equation}
The reflection amplitude $R_s(\alpha)$ was found recently in\cite{Fateev:2001mj}. Completely degenerate fields which contain  $n-1$ null-vectors in their Verma moduli are parametrized by two highest weights $\Omega_{1},\Omega_{2}$ of the Lie algebra $sl(n)$ and correspond to $\alpha=-b\Omega_{1}-\Omega_{2}/b$ \cite{Fateev:1987zh}. In particular, it follows from the definition of the fields $W_{j}(z)$ \eqref{WCurrents} that in the classical case($q\rightarrow 1/b$) the field $V_{-b\omega _{1}}$ satisfies the differential equation of the order $n$ \cite{Fateev:1987zh}
\begin{equation}\label{DiffEqn}
\sum_{k=0}^{n}W_{n-k}(z)(b^{-1}\partial)^{k}V_{-b\omega _{1}}=0. 
\end{equation}
One can expect that in the quantum case such a holomorphic equation \eqref{DiffEqn} should remain in a sense. It was done for the $sl(2)$ Toda or the Liouville Field Theory in \cite{Belavin:1984vu}. The precise statement is the following: all four-point correlation functions that contain at least one degenerate field satisfy  an ordinary differential equation. In particular, the correlation function with the degenerate field $V_{-mb/2}$ satisfies the ordinary differential equation of order $m+1$. The solution to this equation which gives four point correlation function with one degenerate field $V_{-mb/2}$ can be written in terms of  functions $G_{m}^{(c,d,g)}(z)$ defined in \cite{Dotsenko:1984nm,Dotsenko:1984ad}, 
\footnote{Here we use common Liouville normalization $\Delta(a)=a(2Q-a)$ and $2Q=b+1/b$.}
\begin{multline*}
\langle V_{-\frac{mb}{2}}(z,\bar{z})V_{a_{1}}(0)
V_{a_2}(1)V_{a_{3}}(\infty)\rangle\sim\\
\sim|z|^{2mba_{1}}|1-z|^{2mba_{2}}G_{m}^{(c,d,g)}(z),
\end{multline*}
where
\begin{eqnarray*}
 c&=&b(a_{2}+a_{3}-a_{1}-2Q+m/2), \\
 d&=&b(a_{1}+a_{3}-a_{2}-2Q+m/2), \\
 g&=&-b(a_{1}+a_{3}+a_{2}-2Q-m/2)
\end{eqnarray*}
and
\begin{multline}
G_{m}^{(c,d,g)}=\\=
\int\prod_{i=1}^{m}d^{2}t_{i}|t_{i}|^{2c}
|t_{i}-1|^{2d}|t_{i}-z|^{2g}\prod_{i<j}|t_i-t_j|^{-4b^2}. 
\end{multline}
We will show below that in the general $sl(n)$ case such a statement is not valid and the additional restrictions on $\alpha_i$ should be imposed.

Let us assume now that $n$ is general and consider a four-point correlation function
\begin{equation}
\Psi(z)=
\langle V_{-b\omega_1}(z)V_{\alpha_1}(z_1)
V_{\alpha_2}(z_2)V_{\alpha_3}(z_3)\rangle.
\end{equation}
Here $V_{\alpha_k}(z_k)$ are some general primary fields and we have omitted their $\bar{z}$ dependence for simplicity. Let such a function satisfies the order differential equation of the order  $n$
\begin{equation}\label{DiffEqnQ}
\left(-\partial^n+P(z)\partial^{n-1}+\dots\right)\Psi(z)=0.
\end{equation}
Then it should have the following set of the canonical solutions
\begin{equation}\label{Rho}
\Psi_k^j(z)=(z-z_k)^{\rho_k^j}\left(1+\dots\right)
\end{equation}
where the numbers $\rho_k^j=\Delta(\alpha_k-bh_j)-\Delta(\alpha_k)-\Delta(-b\omega_1)$ are known from the OPE \cite{Fateev:1987zh}
\begin{multline*}
V_{-b\omega_1}(z,\bar{z})V_{\alpha_k}(0,0)=\\=\sum_{j=1}^{n}
C_{-b\omega_1,\,\alpha_k}^{\alpha_k-bh_j}\left(|z|^{2\rho_k^j}
V_{\alpha_k-bh_j}(0,0)+\dots\right).
\end{multline*}
Here $C_{-b\omega_1,\,\alpha_k}^{\alpha_k-bh_k}$ are the structure constants of the operator algebra and $\dots$ mean the contribution of the descendants fields. Using \eqref{Rho} one obtains the main asymptotic of $P(z)$   
\begin{equation*}
P(z)=\sum_{k=1}^{3}\frac{p_k}{z-z_k}+\dots
\end{equation*}
with 
$$
p_k=\sum_{j=1}^n\rho_k^j-\frac{1}{2}n(n-1).
$$ 
On the other hand, $\Psi(z)$ should satisfy projective Ward identities. It means that its $z-$dependence is very special
\begin{equation}\label{PWI}
\Psi(z)\sim\frac{\Psi(x)}{(z-z_2)^{2\Delta(-b\omega_1)}},
\quad x=\frac{z_{23}}{z_{13}}\frac{z-z_1}{z-z_2}.
\end{equation} 
The equations \eqref{DiffEqnQ} and \eqref{PWI} are compatible if
\begin{equation}\label{nRelat}
\sum_{k=1}^{3} p_k=(n^2-1)b^2.
\end{equation} 
In the case under consideration(all $\alpha_i$ are general) we find that
\begin{equation*}
p_k=\frac{1}{2}n(n-1)b^2\quad k=1,2,3.
\end{equation*} 
Unfortunately \eqref{nRelat} is satisfied only for $n=2$. And this is the reason why the differential equation in the Liouville Field Theory exists.

One can notice that this difficulty is solved if we suppose that $\alpha_3=\varkappa\omega_{n-1}$. Then only two fields  appear in the OPE\footnote{It follows from the explicit formula for $C_{-b\omega_1,\,\alpha}^{\alpha-bh_k}$ \eqref{StrConst}.}
\begin{equation*}
V_{-b\omega_1}V_{\varkappa\omega_{n-1}}=
\left[V_{\varkappa\omega_{n-1}-bh_1}\right]+\left[V_{\varkappa\omega_{n-1}-bh_n}\right].
\end{equation*} 
Now $p_3=(n-1)b^2$ and $p_1,p_2$ remain unchanged.
It compensates the balance in the sum $\sum p_k$ and \eqref{nRelat} is satisfied. In this case one can write down the differential equation explicitly. Namely, defining a new function
\begin{multline}\label{CorrFunct}
   \langle V_{-b\omega_1}(x)V_{\alpha_1}(0)
   V_{\alpha_2}(\infty)V_{\varkappa\omega_{n-1}}(1)\rangle=\\=
  |x|^{2b(\alpha_1,h_1)}|1-x|^{2\frac{b\varkappa}{n}}G(x,\bar{x}),
\end{multline}
where $G(x,\bar{x})$ satisfies generalized Pochgammer hypergeometric equation
\begin{multline}\label{DiffEqnN}
  \left[x(x\partial+A_1)\dots(x\partial+A_n)-\right.\\\left.
  -(x\partial+B_1-1)\dots(x\partial+B_{n-1}-1)x\partial\right]
  G(x,\bar{x})=0
\end{multline}
with
\begin{equation*}
  A_k=\frac{b\varkappa}{n}-\frac{n-1}{n}b^2+b(\alpha_1-Q,h_1)+
  b(\alpha_2-Q,h_k)
\end{equation*}
and
\begin{equation*}
  B_k=1+b(\alpha_1-Q,h_1-h_{k+1}).
\end{equation*}
Of course, the same equation is valid if we change $x\rightarrow\bar{x}$. The simultaneous single-valued solution to the both equations is unique up to a multiplicative constant and has a simple integral representation
\begin{equation}\label{IntRepr}
G(x,\bar{x})=\int
\prod\limits_{i=1}^{n-1}d^{2}t_{i}\left\vert t_{i}\right\vert
^{2c_i}\left\vert t_{i}-t_{i+1}\right\vert ^{2d_i}\left\vert
t_{1}-x\right\vert ^{2g}, 
\end{equation}
where $t_n=1$ and
\begin{equation*}
c_i=A_i-B_i,\quad
d_i=B_i-A_{i+1}-1,\quad
g=-A_1.
\end{equation*}

Now let us explore \eqref{IntRepr} to obtain the  three-point correlation function of the primary fields
\begin{multline}
C(\alpha_1,\alpha_2,\alpha_3)=|z_{12}|^{2\gamma_{12}}
|z_{13}|^{2\gamma_{13}}|z_{23}|^{2\gamma_{23}}\times\\\times
\langle V_{\alpha_1}(z_1,\bar{z}_1)V_{\alpha_2}(z_2,\bar{z}_2)
V_{\alpha_3}(z_3,\bar{z}_3)\rangle,
\end{multline}
where $\gamma_{ij}$ can be found in \cite{Belavin:1984vu}.
We can rewrite \eqref{CorrFunct} using $s-$channel OPE decomposition 
\begin{multline}\label{Sfusion}
    \langle V_{-b\omega_1}(x)V_{\alpha_1}(0)
   V_{\alpha_2}(\infty)V_{\varkappa\omega_{n-1}}(1)\rangle=\\=
    \sum_{j=1}^nC_{-b\omega_1,\,\alpha_1}^{\alpha_1-bh_j}
    C(\alpha_1-bh_j,\,\alpha_2,\,\varkappa\omega_{n-1})
    \left|\Psi_j(x)\right|^2,
\end{multline}
with $\Psi_j=x^{b(\alpha_1,h_1)}(1-x)^{\frac{b\varkappa}{n}}G_j(x)$ . The functions $G_j(x)$ are expressed in terms of generalized hypergeometric functions of the type $(n,n-1)$ \cite{H.Bateman}
\begin{multline*}
    F\left(\genfrac{}{}{0pt}{1}{A_1\:\dots\:A_n}{B_1\:
    \dots\:B_{n-1}}\biggl|x\right)=1+\frac{A_1\dots A_n}
    {B_1\dots B_{n-1}}x+\\+
    \frac{A_1(A_1+1)\dots A_n(A_n+1)}
    {B_1(B_1+1)\dots B_{n-1}(B_{n-1}+1)}\frac{x^2}{2}+\dots,
\end{multline*} 
as 
\begin{eqnarray*}
    G_1(x)&=&F\left(\genfrac{}{}{0pt}{1}{A_1\:\dots\:A_n}{B_1\:
    \dots\:B_{n-1}}\biggl|x\right),\\
    G_2(x)&=&x^{1-B_1}F\left(\genfrac{}{}{0pt}{1}{1-B_1+A_1\:
    \dots\:1-B_1+A_n}{2-B_1\:\dots\:1-B_1+B_{n-1}}\biggl|x\right),\\
    &&\dots\dots\dots\dots\\ 
    &&\dots\dots\dots\dots\\
    G_n(x)&=&x^{1-B_{n-1}}F\left(\genfrac{}{}{0pt}{1}
    {1-B_{n-1}+A_1\:\dots\:1-B_{n-1}+A_n}
    {1-B_{n-1}+B_1\:\dots\:2-B_{n-1}}\biggl|x\right).\\
\end{eqnarray*}
The ratio of the coefficients before $|\Psi_j(x)|^2$ in \eqref{Sfusion} 
can be found from the integral representation \eqref{IntRepr} 
explicitly in terms of the $\gamma$-functions,  
$\gamma(x)=\frac{\Gamma(x)}{\Gamma(1-x)}$
\begin{multline}\label{RatioOfC}
   \frac{C_{-b\omega_1,\,\alpha_1}^{\alpha_1-bh_1}
    C(\alpha_1-bh_1,\,\alpha_2,\,\varkappa\omega_{n-1})}
    {C_{-b\omega_1,\,\alpha_1}^{\alpha_1-bh_k}
    C(\alpha_1-bh_k,\,\alpha_2,\,\varkappa\omega_{n -1})}=\\
    =\prod_{j=1}^{n}\frac{\gamma(A_j)\gamma(B_{k-1}-A_j)}
    {\gamma(B_j)\gamma(B_{k-1}-B_j)}.
\end{multline}
Here we have set $B_0=B_n=1$. The structure constants 
$C_{-b\omega_1,\,\alpha_1}^{\alpha_1-bh_k}$ admit the free-field 
representation \cite{Fateev:2000ik}
\begin{multline}\label{FreeField}
C_{-b\omega_1,\,\alpha_1}^{\alpha_1-bh_k}=(-\mu)^{k-1}\\
\int\langle V_{-b\omega_1}(0)V_{\alpha_1}(1)
V_{2Q-\alpha_1+bh_k}(\infty)\prod_{i=1}^{k-1}
V_{be_i}(z_i)d^2z_i\rangle_{\small{0}}.
\end{multline}
The expectation value in \eqref{FreeField} is taken using the Wick rules in the theory of a free massless scalar field.
This integral can be calculated explicitly
\begin{multline}\label{StrConst}
    C_{-b\omega_1,\,\alpha_1}^{\alpha_1-bh_k}=
    \left(-\frac{\pi\mu}{\gamma(-b^2)}\right)^{k-1}\times\\\times
    \prod_{i=1}^{k-1}
    \frac{\gamma(b(\alpha_1-Q,h_i-h_k))}
    {\gamma(1+b^2+b(\alpha_1-Q,h_i-h_k))}.
\end{multline}
The equation \eqref{RatioOfC} together with \eqref{StrConst} give us the set of functional relations for the three-point function $C(\alpha_1,\alpha_2,\varkappa\omega_{n-1})$. There is another "dual" set of equations with $b$ replaced by $1/b$. One can readily solve them
\begin{multline}\label{C}
    C(\alpha_1,\alpha_2,\varkappa\omega_{n-1})=
    \left[\pi\mu\gamma(b^2)b^{2-2b^2}\right]^
    {\frac{(2Q-\sum\alpha_i,\rho)}{b}}\times\\
    \frac{(\Upsilon_0)^{n-1}\Upsilon(\varkappa)
    \prod\limits_{e>0}\Upsilon((Q-\alpha_1,e))\Upsilon((Q-\alpha_2,e))}
    {\prod\limits_{ij}\Upsilon\left(\frac{\varkappa}{n}+(\alpha_1-Q,h_i)
    +(\alpha_2-Q,h_j)\right)},
\end{multline}
here $\Upsilon(x)$ is entire selfdual function defined in 
\cite{Zamolodchikov:1995aa}, which satisfies the relations 
\begin{eqnarray*}
\Upsilon(x+b)&=&\gamma(bx)b^{1-2bx}\Upsilon(x),\\
\Upsilon(x+1/b)&=&\gamma(x/b)b^{2x/b-1}\Upsilon(x).
\end{eqnarray*} 
with normalization condition $\Upsilon(1/2(b+1/b))=1$, and
$$
\Upsilon_0=\frac{d\Upsilon(x)}{dx}\biggl|_{x=0}.
$$
Such a function has a semisimple limit \cite{Thorn:2002am}
\begin{equation}
\Upsilon(by)\longrightarrow\frac{\Upsilon_0b^{1-y}}{\Gamma(y)}
\quad\text{as}\quad b\longrightarrow 0.
\end{equation} 
In the numerator of \eqref{C} the product goes over all positive roots 
of the $sl(n)$ and in $(2Q-\sum\alpha_i,\rho)$ the sum includes 
$\alpha_1,\alpha_2$ and $\varkappa\omega_{n-1}$. 

We propose \eqref{C} as an exact three-point function in Conformal Toda Field Theory. Of course, the same as above is true if we consider 
$\alpha_3=\varkappa\omega_1$. The answer would be the same as 
\eqref{C}, but one should use the weights of the fundamental 
representation $\pi _{n-1}$ instead of $\pi _{1}$.

Several simple checks of \eqref{C} can be made. In particular the reflection with respect to the Weyl group $\mathcal{W}$:
$\alpha_1\rightarrow Q+s(\alpha_1-Q):$  $s\in\mathcal{W}$ gives the reflection amplitude
\begin{equation*}
C(Q+s(\alpha_1-Q),\alpha_2,\varkappa\omega_{n-1})=R_s(\alpha_1)
C(\alpha_1,\alpha_2,\varkappa\omega_{n-1}),
\end{equation*}
with
\begin{equation}\label{R}
R_s(\alpha)=\frac{A_s(\alpha)}{A(\alpha)},
\end{equation}
where
\begin{multline*}
A(\alpha)=\left(\pi\mu\gamma(b^2)\right)^
{\frac{(\alpha-Q,\rho)}{b}}\times\\\times
\prod_{e>0}\Gamma(1-b(\alpha-Q,e))\Gamma(1-(\alpha-Q,e)/b).
\end{multline*}
The result \eqref{R} was obtained previously in \cite{Fateev:2001mj}.

If the parameters in \eqref{C} satisfy the screening condition
\begin{equation*}
\alpha_1+\alpha_2+\varkappa\omega_{n-1}+
b\sum_{i=1}^{n-1}l_{i}e_{i}=2Q 
\end{equation*}
with some non-negative integer numbers $l_i$, \eqref{C} should have 
a multiple pole of the order $n-1$ with the residue being expressed in 
terms of  a free field integral \cite{Goulian:1990qr}. Such an integral was calculated for  the Liouville case in \cite{Dotsenko:1984nm,Dotsenko:1984ad}. The general $sl(n)$ case was done recently \cite{Fateev} and the result agrees with \eqref{C}.

It is interesting to consider the semiclassical limit $b\rightarrow 0$. In this limit in the Hamiltonian picture, associated with radial quantization, we take into account only the zero modes dynamics
(minisuperspace approach) \cite{Seiberg:1990eb,Moore:1991ir}. In this approximation the state created by operator $V_{Q+iP_j}$ corresponds to wave function
\begin{equation*}
V_{Q+iP_j}\rightarrow \Psi _{P_j}(x),
\end{equation*} 
where $x$ is a zero mode of field $\varphi$. The function $\Psi _{P}(x)$ ($sl(n)$ Whittaker function) satisfy Scr\"odinger equation
\begin{equation}
\left(-\nabla_{x}^{2}+8\pi\mu\sum_{i=1}^{n-1} e^{b(e_{i}x)}\right)
\Psi _{P}(x)=P^{2}\Psi _{P}(x),
\end{equation}
and in the region $(e_i,x)<0$ (Weyl chamber) posses the asymptotic
\begin{equation*}
\Psi _{P}(x)\sim\exp (i(P,x))+\sum_{s\in\mathcal{W}}S_{s}(P)\exp (i(s(P),x)),
\end{equation*}
where the sum runs over all elements of the Weyl group $\mathcal{W}$ besides identical and the coefficients $S_{s}(P)$ are known 
exactly \cite{Olshanetsky:1981dk}\footnote{Note also, that $S_{s}(P)$  can be obtained from the reflection amplitude $R_s(Q+iP)$ in semiclassical limit $b\rightarrow 0$.}.
The minisuperspace approximation is valid if $P_j/b$ are fixed. If
we take $\varkappa=ibs$ and $P_i=ibp_i$ then the semiclassical limit 
of the three-point correlation function should be given by the integral
\begin{multline}\label{mspLimit}
C(Q+ibp_1,Q+ibp_2,bs\omega_{n-1})\longrightarrow \\
\longrightarrow\int d\vec{x}\,\Psi _{bp_{1}}(x)\Psi _{bp_{2}}(x)e^
{ibs(\omega_{n-1},x)}. 
\end{multline}
The theory of the $sl(n)$ Whittaker functions has some long history. 
In particular an explicit integral representation for these functions
exists \cite{Stade:1990,Kharchev:1999bh,Kharchev:2000ug}. Resent progress was done in \cite{Stade:2002}, where the integral in the right hand side of \eqref{mspLimit} was calculated
\begin{multline}\label{StadeFormula}
    \int d\vec{x}\,\Psi _{bp_{1}}(x)\Psi_{bp_{2}}(x)
    e^{ibs(\omega_{n-1},x)}=\\=
    \frac{1}{b^{n-1}}
    \left(\frac{\pi\mu}{b^2}\right)^
    {-i(s\frac{(n-1)}{2}+(p_1+p_2,\rho))}
    \times\\\times
    \frac{\prod\limits_{ij}\Gamma\left(\frac{is}{n}+i(p_1,h_i)
    +i(p_2,h_j)\right)}{\Gamma(is)\prod\limits_{e>0}
    \Gamma(-i(p_1,e))\Gamma(-i(p_2,e))}.
\end{multline}
We note that the result \eqref{StadeFormula} coincides exactly with the 
corresponding limit of the three-point function \eqref{C}. Unfortunately, the integral \eqref{StadeFormula} with arbitrary parameter of the Fourier transform is more complicated object and its analytical expression is still unknown. In quantum case this integral corresponds to the semiclassical limit of general three-point function(with all $\alpha_i$ being arbitrary). We suppose to investigate the general situation in more details in future publications. 

The authors are grateful to D.Lebedev for explanation of some recent results from the theory of the Whittaker functions and to E.Stade for bringing the paper \cite{Stade:2002} to our attention. A.L. would like to thank A.Belavin, B.Feigin, M.Lashkevich and Ya.Pugai for useful discussions and suggestions. This work was supported, in part, by EU under contract EUCLID HRPN-CT-2002-00325, by INTAS under the grant INTAS-OPEN-03-51-3350, by Russian Foundation for Basic Research under the grant RBRF 04-02-1602 and by Russian Ministry of Science and Technology under the Scientific Schools grant 2044.2003.2.  A.L. thanks also the Forschungszentrum Juelich for financial support within the framework of the Landau Program and the "Dynasty" foundation. 

\end{document}